\documentclass[twocolumn,aps, prb, showpacs, preprintnumbers, amsmath, amssymb, a4paper, superscriptaddress]{revtex4-1}

\usepackage{graphicx}
\usepackage{dcolumn}
\usepackage{bm}
\usepackage{color}

\begin{document}
\newcommand{\Arg}[1]{\mbox{Arg}\left[#1\right]}
\newcommand{\bb}{\mathbf}
\newcommand{\braopket}[3]{\left \langle #1\right| \hat #2 \left|#3 \right \rangle}
\newcommand{\braket}[2]{\langle #1|#2\rangle}
\newcommand{\be}{\[}
\newcommand{\br}{\vspace{4mm}}
\newcommand{\bra}[1]{\langle #1|}
\newcommand{\braketbraket}[4]{\langle #1|#2\rangle\langle #3|#4\rangle}
\newcommand{\braop}[2]{\langle #1| \hat #2}
\newcommand{\dd}[1]{ \! \! \!  \mbox{d}#1\ }
\newcommand{\DD}[2]{\frac{\! \! \! \mbox d}{\mbox d #1}#2}
\renewcommand{\det}[1]{\mbox{det}\left(#1\right)}
\newcommand{\ee}{\]} 
\newcommand{\eg}{\textbf{\\  Example: \ \ \ }}
\newcommand{\Imag}[1]{\mbox{Im}\left(#1\right)}
\newcommand{\ket}[1]{|#1\rangle}
\newcommand{\ketbra}[2]{|#1\rangle \langle #2|}
\newcommand{\kp}{\arccos(\frac{\omega - \epsilon}{2t})}
\newcommand{\ldos}{\mbox{L.D.O.S.}}
\renewcommand{\log}[1]{\mbox{log}\left(#1\right)}
\newcommand{\Log}{\mbox{log}}
\newcommand{\Modsq}[1]{\left| #1\right|^2}
\newcommand{\nb}{\textbf{Note: \ \ \ }}
\newcommand{\op}[1]{\hat {#1}}
\newcommand{\opket}[2]{\hat #1 | #2 \rangle}
\newcommand{\occ}{\mbox{Occ. Num.}}
\newcommand{\Real}[1]{\mbox{Re}\left(#1\right)}
\newcommand{\so}{\Rightarrow}
\newcommand{\sol}{\textbf{Solution: \ \ \ }}
\newcommand{\thetafn}[1]{\  \! \theta \left(#1\right)}
\newcommand{\tin}{\int_{-\infty}^{+\infty}\! \! \!\!\!\!\!}
\newcommand{\Tr}[1]{\mbox{Tr}\left(#1\right)}
\newcommand{\kb}{k_B}
\newcommand{\rad}{\mbox{ rad}}
\preprint{APS/123-QED}

\title{Electronic transport in graphene nanoribbons with sublattice-asymmetric doping}

\author{Thomas Aktor}
\author{Antti-Pekka Jauho}
\author{Stephen R. Power}
\email{spow@nanotech.dtu.dk}
\affiliation{Center for Nanostructured Graphene (CNG), DTU Nanotech, Department of Micro- and Nanotechnology,
Technical University of Denmark, DK-2800 Kongens Lyngby, Denmark}

\date{\today}

\begin{abstract}
Recent experimental findings and theoretical predictions suggest that nitrogen-doped CVD-grown graphene may give rise to electronic band gaps due to impurity distributions which favour segregation on a single sublattice.
Here we demonstrate theoretically that such distributions lead to more complex behaviour in the presence of edges, where geometry determines whether electrons in the sample view the impurities as a gap-opening average potential or as scatterers.
Zigzag edges give rise to the latter case, and remove the electronic bandgaps predicted in extended graphene samples.
We predict that such behaviour will give rise to leakage near grain boundaries with a similar geometry or in zigzag-edged etched devices.
Furthermore, we examine the formation of one-dimensional metallic channels at interfaces between different sublattice domains, which should be observable experimentally and offer intriguing waveguiding possibilities.

\end{abstract}

                 
\maketitle

\section{Introduction}
\label{intro}
The high Fermi velocity and linear electronic dispersion in graphene appear promising for electronic devices.\cite{neto:graphrmp}
The absence of an intrinsic band gap is a potential stumbling block for many applications.
A range of possibilities are being investigated to redress this.
Many involve geometric constraints in the form of, \emph{e.g.}, finite-width nanoribbons (GNRs)\cite{Son:ribbonenergygaps} or periodic perforations.\cite{Pedersen:GALscaling}
An alternative route is the manipulation of the atomic level structure.
The hexagonal graphene lattice is composed of two intersecting triangular sublattices, $A$ and $B$, shown by hollow and filled symbols respectively in the top panels of Fig \ref{fig_trans-dos}. 
The equivalence of these leads to the gapless band structure.
A sublattice dependent potential opens a band gap and gives mass to the charge carriers.
A possible implementation is to place graphene on a substrate, such as hexagonal boron nitride (hBN), which offers a potential varying on approximately the required length scale\cite{PhysRevB.76.073103}.
However, the potential here is quite weak and lattice mismatches give rise to larger scale Moir\'e features.\cite{PhysRevB.89.201404, NatPhy-GonhBN, NatComm-GonhBN}

Recent experiments suggest another route to breaking sublattice equivalence.
Nitrogen-doped graphene grown by chemical vapour deposition (CVD) can show unusual distributions of substitutional N atoms.
Large domains are found with N atoms primarily occupying a single sublattice.\cite{zhao2011visualizing, lv2012nitrogen, zhao2013local, zabet2014segregation, lawlor2014sublattice}
This behaviour depends on growth conditions and theoretical works suggest possible mechanisms including preferential impurity positioning relative to edges during growth\cite{deretzis2014origin} and inter-impurity interactions in disordered ensembles.\cite{lawlor2013friedel, lawlor2014sublatticeb}
Subsequent studies of N-doped graphene treated by high-temperature annealing\cite{telychko2014achieving}, and of graphene decorated by hydrogen adatoms,\cite{lin2015direct} suggest that asymmetric distributions may also arise in other scenarios.
Such doping leads to different average potentials on each sublattice and is equivalent to introducing an effective mass term.
Extended graphene sheets with sublattice-asymmetric impurity distributions are predicted to display electronic and transport band gaps, and electron-hole asymmetry in their conductivity.  \cite{PhysRevB.77.115109, PhysRevLett.105.086802, PhysRevLett.105.266803, abanin2010spatial, lherbier2013electronic} 

In this work we focus on nanoribbons with sublattice asymmetric doping.
This is motivated both by the possibility of etching\cite{Zhang2012-h-etching} and transferring\cite{Ruoff2009transfer} devices from doped graphene sheets and by the need to understand the interplay between the effective mass term introduced by such doping and effects induced by symmetry breaking edges.
This is important since CVD-grown graphene contains extended edge-like defects in the form of grain boundaries, \cite{lahiri2010extended, huang2011grains, ZettlGGBmapping, Yazyev2014} unlike bottom-up approaches which may allow synthesis of more precise geometries.\cite{cai2010atomically}
We are further motivated by the strong dependence of GNR transport on edge geometry and impurity distribution\cite{li:ribbonedgedefects, mucciolo:graphenetransportgaps, rigo:Nidopedribbons, biel:ribbondoping, Dietl2009, PhysRevB.83.045414, Cruz:subdopedribbons, PhysRevB.83.205125, Chang3NN_pores, owens2013electronic, botello2013modeling, Orlof2013, PhysRevB.89.195406, PhysRevB.92.014405} and by sublattice dependent features in carbon nanotubes.\cite{Thygesen2008, james:CNTs}
We consider both armchair (AGNR) and zigzag (ZGNR) edged ribbons, noting the inbuilt sublattice asymmetry of ZGNRs due to sites along one edge belonging to one sublattice. 
Similar behaviour to bulk graphene is found for AGNRs - namely reliable electronic and transport band gaps consistent with an average mass term model. 
For ZGNRs, only a suppression of transmission is found in the expected gap region and it is not accompanied by a vanishing density of states (DOS).
In particular, strong finite DOS clusters remain along one ZGNR edge.
This is related to the position dependence of simple impurity bound states near zigzag edges and is captured within a coherent potential approximation (CPA) model.
Finally, we investigate interfaces between different sublattice domains and predict that these should give rise to robust one-dimensional metallic wires embedded within the gapped system, and which should have features detectable by scanning tunneling microscopy (STM).

\section{Models}
\label{section_model}

The electronic structure of graphene is well-described by a nearest-neighbour tight-binding Hamiltonian
with a hopping integral $t = -2.7\mathrm{eV}$.
The use of this model is validated in Appendix \ref{appendix3NN}, where  key features from our results are reproduced using a higher-order model, 
We take $|t|$ as the unit of energy and include substitutional N dopants by a change of onsite energy $\Delta = - |t|$.
More accurate parameterisations can be achieved\cite{lherbier2013electronic, Pedersen-SCdoping, Pedersen-edgedoping} but the qualitative behaviour described here is reasonably independent of impurity species or parameterisation. 
We will discuss the change in carrier density induced by such dopants at the end of Section \ref{sec_results} below.
A general band dispersion is given by
\begin{equation}
\epsilon_\pm (\mathbf{k})  =  \frac{1}{2}\left(\epsilon_A + \epsilon_B\right) \pm \frac{1}{2} \sqrt{(\epsilon_A - \epsilon_B)^2 + 4 t^2 \left| f (\mathbf{k}) \right|^2}\,
\label{bands_eqn}
\end{equation}
where $\epsilon_A$ ($\epsilon_B$) is the potential on the $A$ ($B$) sublattice and $f (\mathbf{k})$ is a term arising from the sum of Bloch phases over neighbouring sites.
For pristine graphene, $\epsilon_A = \epsilon_B = 0.0 $, and so $\epsilon_\pm (\mathbf{k}) = \pm t \left| f (\mathbf{k}) \right|$, which is gapless near $E=0$. 
Uniformly breaking the sublattice symmetry, by setting $\epsilon_A \ne \epsilon_B$, has three effects on the bandstructure: i) a bandcentre shift of $\frac{\epsilon_A + \epsilon_B}{2}$, ii) a direct band gap of magnitude $|\epsilon_A - \epsilon_B|$ at the Dirac points and iii) the breaking of the band linearity due to the additive constant $(\epsilon_A - \epsilon_B)^2$ in the square root. 
The quantity $\frac{|\epsilon_A - \epsilon_B|}{2}$ is called a \emph{mass} term, and the dispersion of electrons in the gapped systems is no longer linear or massless. 

Transport quantities are calculated using recursive Green's function (GF) techniques\cite{Lewenkopf2013}. 
Semi-infinite leads are constructed using an efficient decimation procedure\cite{Sancho-Rubio} and the zero-temperature conductance is given by\cite{DattaBook} $G = \frac{2e^2}{h} T$ , where the transmission is calculated from 
$
T(E) = \mathrm{Tr}\big[ \mathbf{G}^r \mathbf{\Gamma}_R  \mathbf{G}^a  \mathbf{\Gamma}_L\big], 
$
where $ \mathbf{\Gamma}_{i}(E)$ ($i=L,R$) are the level width matrices and $ \mathbf{G}^{r/a}(E)$ is the retarded/advanced Green's function of the device region. 
A configurational average is taken for disordered systems to discern the overall trends.
We also examine the local density of states (LDOS), which at site $i$ is given by
$
\rho_{i} (E_F)= - \frac{1}{\pi} \mathrm{Im} \big[G^r_{ii} (E_F)\big]
$.
The GFs required here involve a double sweep through the device region.\cite{Lewenkopf2013}

Effective medium models are used to analyse the configurationally-averaged densities of states. 
The use of the two different models below allows to isolate effects arising from an average disorder-induced potential or mass-term, and the effects of scattering from individual impurities.
Both models employ a 1NN tight-binding description which is perfectly periodic along the ribbon direction.
Onsite energies within the repeated unit cell are determined as described below.
The \emph{virtual crystal approximation} (VCA) ignores scattering effects and simply takes into account the new average potential felt by electrons.
In practise this is done by introducing a self energy to shift onsite energies by $c \Delta$, where $c$ is the doping concentration and $\Delta$ is the shift caused by a single dopant.\cite{elliott_theory_1974}
For sublattice dependent doping, this is generalised so that the self energy is sublattice dependent, $\Sigma_x = c_x \Delta_x$ for $x=A,B$, due to $c_x$ (and/or $\Delta_x$) taking different values on each sublattice.
This new unit cell is then considered part of an infinite perfectly periodic \emph{virtual crystal} allowing us to calculate the Green's function and thus the density of states.
The \emph{coherent potential approximation} (CPA) replaces this potential with a position and energy dependent self energy to include simple scattering effects. 
This self energy is found from the solution of the self-consistent equation
$
\Sigma_x=c_x\Delta_x(1-(\Delta_x-\Sigma_x)G_{\textrm{eff}})^{-1} \,,
$
where $G_{\textrm{eff}}$ is the Green's function of the new effective medium. \cite{elliott_theory_1974, soven_contribution_1969}
Periodicity of the effective medium along the ribbon direction can again be used to quickly calculate the Green's function and density of states.
It can be shown that the CPA includes simple scattering effects, but neglects higher-order scattering terms.
In this way features appearing in the CPA, but not the VCA, arise due to the scattering effects beyond an average potential but below higher-order cluster effects such as localization.

%

\section{Results and Discussion}
\label{sec_results}

\begin{figure}
	\centering
	\includegraphics[width =0.45\textwidth]{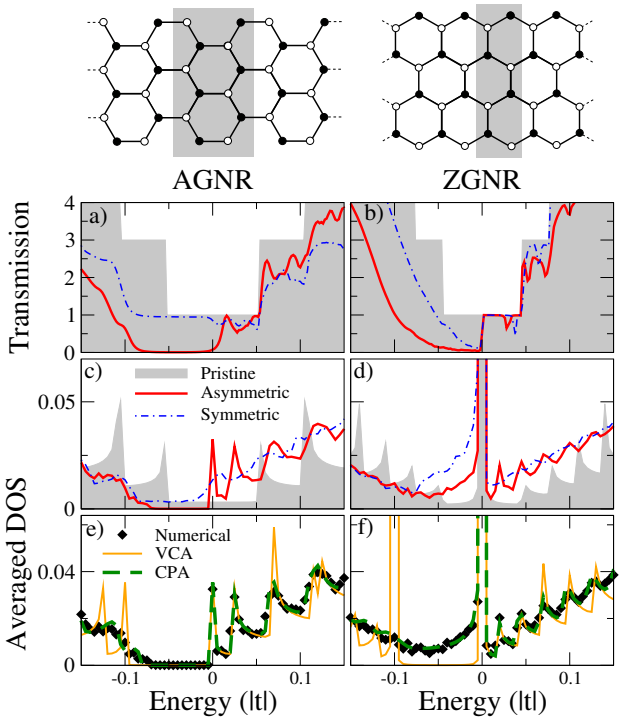}
	\caption{(Top) Schematics of a 6-AGNR  and 4-ZGNR, with the unit cells shown by the shaded areas and the A (B) sublattice sites by hollow (filled) symbols. The index counts the dimer lines or zigzag chains across the ribbon. 		
		Remaining panels show results for a 101-AGNR (left) and a 100-ZGNR (right).
		a), b) show the (averaged) transmission through pristine systems (grey shading) and also systems with $40$ unit cells of sublattice asymmetric (solid red lines) and sublattice symmetric disorder (blue dashed-dotted line). c), d) show the numerically averaged DOS of longer systems with corresponding disorder profiles. e), f) show the numerically averaged DOS for the fully asymmetric case (black symbols) compared to VCA (orange) and CPA (green, dashed) model calculations. The concentration of N atoms for all disordered cases is $5\%$.}
	\label{fig_trans-dos}
\end{figure}

We first calculate the transmission through both GNR types for two disorder types -- a completely random distribution of impurities over all sites (\emph{symmetric}) or a distribution confined to only one sublattice (\emph{asymmetric}). 
Fig. \ref{fig_trans-dos} shows transmissions through a) 101-AGNR (width $\sim12$ nm)  and b) 100-ZGNR (width $\sim21$ nm) systems.
In the absence of disorder, these ribbons are both metallic within the nearest-neighbour tight-binding approximation. 
Results for the initially semiconducting 100-AGNR are shown in Appendix \ref{appendix3NN}.

The conductance of the pristine systems is shown by the grey shaded areas and the averaged asymmetrically (symmetrically) doped systems by solid red (dashed blue) lines.
Configurational averages over $100$ instances of disorder through device regions $40$ unit cells long ($17$ nm for AGNR, $10$ nm for ZGNR) are shown.
Impurity concentrations are $c_A = 0.05, c_B=0.05$  ($c_A = 0.1, c_B=0.0$) for the symmetric (asymmetric) cases, where $c_{A/B}$ is the concentration on a given sublattice.
Note that the asymmetric case corresponds to a random replacement of $10\%$ of sublattice A carbon atoms with nitrogen atoms within the disordered region, for a total nitrogen concentration of $5\%$ as the B sublattice is unaltered.
The total concentration of nitrogen is thus the same for both cases.

For AGNRs, asymmetric disorder opens a band gap with sharp edges on the hole side of the spectrum, in contrast to symmetric disorder where very little transmission suppression is seen.
The persistence of the $T=1$ plateau in the symmetric case has been observed previously \cite{Dietl2009}.
In general, AGNRs are more sensitive to edge disorders than the bulk substitutional disorder considered here.\cite{mucciolo:graphenetransportgaps, Orlof2013}
The transport gap for asymmetric doping has a corresponding electronic band gap, clearly visible in the averaged DOS plot in c). 
This shows an average over the central 800 cells of a disordered region with total length 1000 unit cells.
The appearance of this band gap is consistent with the results for similarly doped extended graphene sheets\cite{lherbier2013electronic}.
A comparison of the numerically averaged DOS to results from the VCA and CPA models is shown for the fully asymmetric case in Fig. \ref{fig_trans-dos}e). 
Good agreement between the VCA and numerical results is seen within the gap and on the electron side, while poor agreement is seen on the hole side.
The VCA also overestimates the bandgap, which is somewhat smaller than the value $c_A \Delta = 0.1 |t|$ given by a uniform mass term.
These discrepancies are almost entirely corrected by the CPA, where excellent agreement is seen over the entire energy range.

The accuracy of the VCA at gap and electron-side energies suggests that the main effect of disorder here is not scattering, but rather an averaged potential landscape with a sublattice dependent mass term. 
The unimportance of scattering effects here is also apparent in the transmission shown in Fig. \ref{fig_trans-dos}a), where the asymmetric disorder only induces minor quenching of transmission at these energies.
Conversely, the failure of the VCA and success of the CPA on the hole side suggest that scattering plays a more important role here.
This is further evidenced by the hole-side transmission, which is significantly reduced relative to the pristine case and has its plateau features almost completely smeared out.
This electron-hole asymmetry is consistent with results in graphene sheets, where reduced mobility on the hole side is associated with a pseudospin polarisation giving a higher occupation of the undoped (doped) sublattice on the electron (hole) side. \cite{lherbier2013electronic}
We have confirmed that this feature is also present in the AGNR case by examining the sublattice dependent averaged DOS.
The reduced gap size compared to the VCA prediction is in line with a sublinear gap dependence found in graphene sheets. \cite{PhysRevB.77.115109, lherbier2013electronic}
We have varied the concentration and find agreement with the $E_G \sim c_A^{0.75}$ scaling previously reported.\cite{lherbier2013electronic}

The right-hand side panels of Fig. \ref{fig_trans-dos} show that many of the features discussed above are radically altered for zigzag edged systems.
Transmission suppression is observed in the gap region for asymmetric doping in Fig. \ref{fig_trans-dos}b), but without sharply defined band gap edges.
Furthermore, a significant DOS is noted throughout the expected bandgap (Fig. \ref{fig_trans-dos}d).
It is thus unsurprising that the VCA (Fig. \ref{fig_trans-dos}f) fails to capture the DOS features at these energies, since this model always gives a bandgap.
However, it does capture the low-energy electron-side behaviour quite well, including the sharp peak at $E=0$.
This peak is associated with states localised on the edge atoms of a ZGNR.
It is doubly-degenerate in pristine ribbons, as the states on each ribbon edge, although belonging to opposite sublattices, are equivalent.
Adding a uniform mass term breaks this degeneracy and the peak splits into two which reside at the bandgap edges, at energies corresponding to the onsites of each sublattice.
This is seen for the VCA result, but the peak at $E=-0.1|t|$, associated with the N-doped sublattice, is absent in the numerical results and only the undoped sublattice peak remains.
The CPA once again restores the features absent within the VCA, suggesting that the finite DOS in the expected band gap is due to scattering processes dominating over a gap-opening average potential.

\begin{figure}
	\centering
	\includegraphics[width =0.45\textwidth]{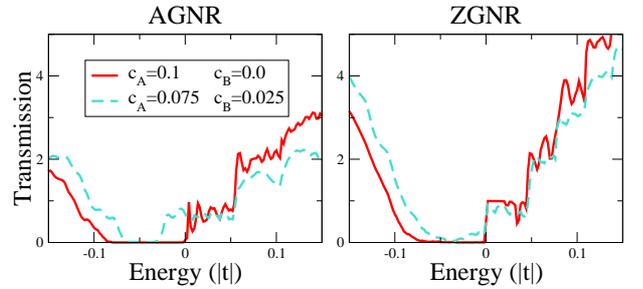}
	\caption{Transmissions for 101-AGNR (left) and 100-ZGNR systems with 80 unit cells of asymmetric disorder. Results are shown for both fully (red, solid) and partially (turquoise, dashed) asymmetric distributions of impurities.}
	\label{fig_7525}
\end{figure}

To verify the robustness of the gap-opening feature, we consider the case of less than perfect sublattice asymmetry. 
Fig \ref{fig_7525} shows the transmissions through systems analogous to those in Fig. \ref{fig_trans-dos}a) and b), but with 75\% of N atoms on sublattice A and 25\% on sublattice B. 
Curves for a fully asymmetric case are shown for comparison.
For partial asymmetry, we note a clear band gap formation for the AGNR case, whereas transmission suppression without a clear band gap is still present for the ZGNR case. 
The AGNR band gap is shifted away from $E=0$, unlike that of the perfectly asymmetric case, as the band centre shift and mass terms entering in Eq. \eqref{bands_eqn} are no longer equal.
Band gap formation at this level of asymmetry is promising for realising such a gap experimentally, as samples with over 90\% asymmetry have been reported.\cite{zabet2014segregation}

\begin{figure}
	\centering
	\includegraphics[width =0.45\textwidth]{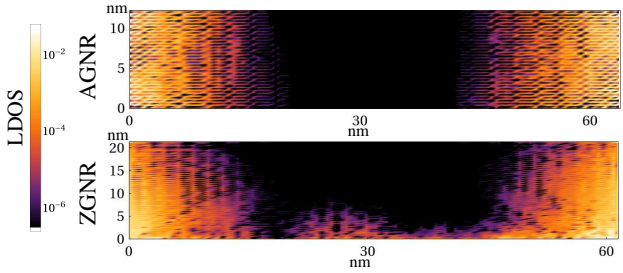}
	\caption{LDOS maps of a disordered 101-AGNR (top) and 100-ZGNR (bottom) at $E=-0.04|t|$. The impurities are entirely on the $A$ sublattice, corresponding to the bottom edge of the ZGNR, where a non-vanishing DOS is evident. }
	\label{fig_GNR-maps}
\end{figure}

To further explore the differences between armchair and zigzag edged geometries, we show LDOS maps for a single, fully-asymmetric disorder configuration of each in Fig \ref{fig_GNR-maps}.
The maps are taken at an energy in the middle of the expected band gap.
The LDOS decays quickly as we move into the disordered region of an AGNR.
This decay is also uniform across the ribbon width.
For the ZGNR, the LDOS vanishes throughout most of the system.
However, large clusters of finite density remain along the bottom edge of the ribbon, which is associated with the doped sublattice.
This suggests an interplay between the doping of a particular sublattice and the proximity of a zigzag edge of the same sublattice.
The reproduction of averaged DOS features within the CPA model suggests that this effect can be explained in terms of single scattering processes, and so we now examine individual N dopants in a ZGNR.

\begin{figure}
	\includegraphics[width =0.45\textwidth]{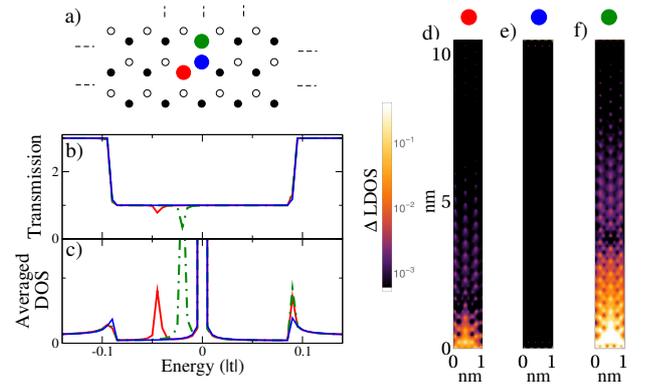}
	\caption{The transmission (b) and averaged DOS (c) for a 50-ZGNR with a single N impurity located at each of the sites shown by the symbol of the same colour in  a). d)-f) map the change in LDOS near the three possible impurity locations, taken at $E=-0.05|t|$ (d) and $E=-0.02|t|$ (e and f).}
	\label{fig_single-imps}
\end{figure}

Fig \ref{fig_single-imps}a shows a few possible sites for a single N atom near the edge of a $50$-ZGNR.
The sites represented by red and green circles are on the edge sublattice, whereas the blue site is not.
Fig \ref{fig_single-imps}b) and c) show that impurity sites on the edge sublattice give rise to conductance dips and corresponding DOS peaks in the low energy window shown here.
These features, associated with anti-resonances formed by the impurity, have been studied previously in GNRs\cite{li:ribbonedgedefects, Orlof2013}.
Symmetry-breaking edges result in a strong position dependence of the anti-resonance energies.
Interestingly, sites near a ZGNR edge and of the same sublattice type can give rise to features at energies within the expected band gap, whereas those on the opposite sublattice (and sites in AGNRs) result in features at energies far outside this window.
In Fig \ref{fig_single-imps} d)-f) the change in LDOS near three of these sites is mapped.
For d) and f), corresponding to sites on the edge sublattice, we choose the DOS peak energy and note a significant triangular region of increased DOS near the impurity locations at the bottom edge.
For the opposite-sublattice impurity site in e), we choose the same energy as d), and note that no such feature is visible and the DOS barely differs from that of a pristine ribbon.
Consequently electrons in this energy range are scattered by impurities located on the same sublattice as the edge, and not by those on the opposite sublattice.
Returning to asymmetrically disordered ZGNRs, we can understand the finite DOS in the expected bandgap (Fig \ref{fig_trans-dos}d) as the average of many single impurity peaks at different energies and corresponding to A-sublattice impurities at different locations near the bottom edge. 
Away from this edge, the density of states vanishes as shown in Fig. \ref{fig_GNR-maps}, because the net effect of the doping here is an average mass term and not scattering from impurity states.
This is confirmed further by examining the position dependence of the CPA self-energy, $\Sigma_A$, which in AGNRs takes a real and quite uniform value slightly smaller than $c_A \Delta$.
This is also true across much of a ZGNR, except near the edge associated with the doped sublattice, where $\Sigma_A$ becomes complex and its real part varies drastically from $c_A \Delta$.
The VCA is unable to explain behaviour near this edge, as the net effect of the doping is no longer an effective mass term.
Increasing the device length will lead to a transport gap as we enter the localization regime.
However, this gap is unrelated to the effective mass term or a DOS gap, and is similar to the behaviour observed for ZGNRs with symmetric doping.

\begin{figure}[t]
	\includegraphics[width =0.48\textwidth]{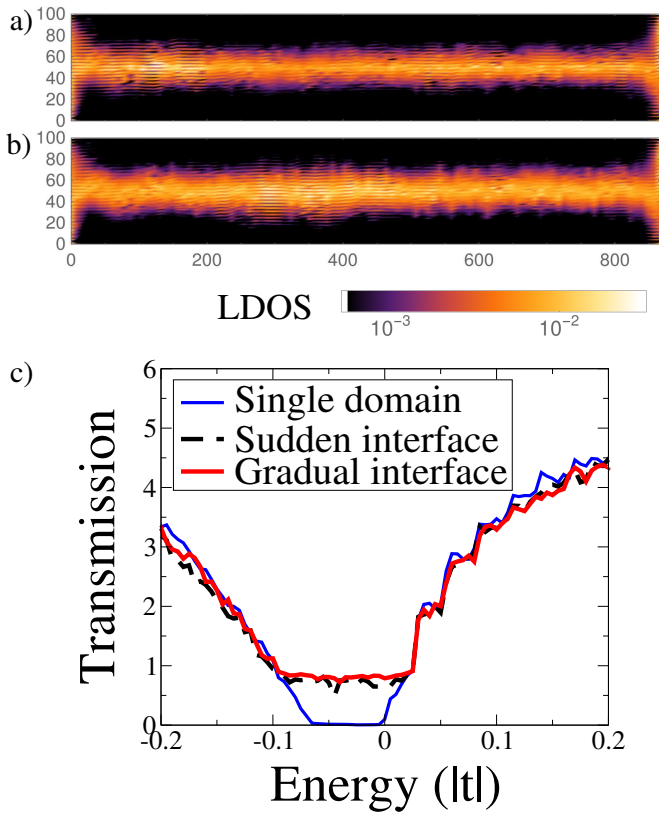}
	\caption{LDOS maps of asymmetrically doped 200-AGNRs with a) sudden or b) gradual sublattice interfaces running along the centre of the ribbon, taken at $E=-0.04|t|$. c) shows the transmissions for these systems compared to one with a single domain.}
	\label{fig_interface}
\end{figure} 

The breakdown of the band gap in asymmetrically doped graphene near a zigzag edge may have interesting consequences beyond GNR devices.
Grain boundaries can have geometries similar to zigzag edges and break the lattice symmetry in the same manner\cite{lahiri2010extended}.
Thus CVD-grown systems may experience leakage near these boundaries.
Another relevant interface is that between neighbouring regions with doping on opposite sublattices. 
These have been mapped experimentally\cite{zabet2014segregation}, and in Fig. \ref{fig_interface} we consider an AGNR with a sublattice interface running parallel to the edge so that only the A (B) sublattice is doped in the bottom (top) of the device.
Near the boundary the average mass term switches sign, closing the band gap and resulting in states confined near the interface \cite{Semenoff2008}. 
This is confirmed in the LDOS maps in Fig. \ref{fig_interface}, shown for systems with both a)\emph{sudden} and b)\emph{gradual} interfaces where the impurity concentration changes linearly from one sublattice to the other over $4$ or $20$ atoms respectively.
In both cases we note a large, finite DOS running along the interface and decaying away from it.
Furthermore this regions acts as a propagating channel as is clear from panel c), where a finite transmission is noted across the band gap region of a single domain device.

Electron doping by nitrogen impurities shifts the Fermi energy, $E_F$, relative to any gap. 
Accessing the gap region experimentally will involve the application of a gate voltage.
While accurate electron-counting can be performed within DFT calculations\cite{lherbier2013electronic} for single impurities or small disordered regions, this is not feasible for the system sizes considered here or in experiment.
Nonetheless, the charge density fluctuation can be 	approximated from $\delta n \sim \tfrac{E_D (c_A + c_B) \rho_C}{2}$, where $E_D \approx 0.4$ is the average doping efficiency of nitrogen in 	GNRs\cite{Pedersen-edgedoping} and $\rho_C$ is the density of lattice sites in graphene.
For $c_A=0.1$, we find $\delta n \sim 7.6 \times 10^{13} \mathrm{\,cm}^{-2}$, just inside the range of the most advanced gating methods. \cite{craciun2011tuneable} $c_A=0.02$ gives a more realistic $\delta n \sim 1.5 \times 10^{13} \mathrm{\,cm}^{-2}$, while yielding $E_G \sim 50\mathrm{-}200 \mathrm{\,meV}$. Gaps from lower concentrations, whilst too small for applications, still allow experimental verification of our results. It is also possible to shift $E_F$ nearer the gap by codoping with a symmetrically distributed \emph{p}-dopant, at the cost of reducing transmission outside the band gap.

\section{Conclusions}
\label{sec_conclude}
Our results highlight the importance of edge geometry in doped graphene nanoribbons.
The band gap predicted for sublattice asymmetrically doped graphene is sensitive to the presence of zigzag edges, where a gap-opening average potential is no longer the dominant effect of disorder.
Instead impurity bound states within the expected band gap, associated with the edge sublattice, lead to a finite DOS and propagation, albeit scattered, along the edge.
A band gap opening, similar to that in graphene sheets, is observed for armchair edges.
The sensitivity of gap opening to edge geometry is relevant beyond ribbon devices.
The majority of samples with sublattice asymmetric disorder are grown by CVD, which gives rise to  edge-like defects in the form of grain boundaries.
Since these can have zigzag-edge like symmetries, we expect similar leakage near grain boundaries in asymmetrically doped polycrystalline graphene sheets.
This may make it difficult to verify experimentally the band gaps predicted for such systems.
Finally, we show the formation of one-dimensional metallic wire behaviour at the interface between two regions with doping on opposite sublattices.
Such interfaces are present in experimental systems, and the features we predict should be observable to STM measurements. 
These channels present waveguiding possibilities as, away from defects or edges, leakage is prevented by the gapped region surrounding them.

\hspace{1cm}

\begin{acknowledgments}
The Center for Nanostructured Graphene (CNG) is sponsored by the Danish National Research Foundation, Project DNRF58.
The authors would like to thank S{\o}ren Schou Gregersen for useful discussions during thr review process.  
\end{acknowledgments}

\appendix

\section{Comparison of 1st and 3rd nearest-neighbour tight-binding results}
\label{appendix3NN}
To check the validity of the first nearest-neighbour tight-binding approximation (1NN) for our systems, we compare the transmissions of pristine and asymmetrically disordered nanoribbons calculated with both this model, and with a more complete third nearest-neighbour (3NN) description of graphene.
We also consider a 100-AGNR which is semiconducting within a 1NN description in the absence of dopants.
The 1NN results are based on the system in  Figs \ref{fig_trans-dos}a) and b), where a constant value of $t =  -2.7\mathrm{eV}$ is used throughout the system to describe the hopping parameter between nearest neighbour sites. 
The 3NN results are calculated using the same relative 2nd and 3rd neighbour hoppings for pristine graphene as in Ref. \onlinecite{lherbier2013electronic}. 
For both models, we use a simple onsite shift of $\Delta = - |t|$ to represent an impurity.
Larger values of $\Delta$, suggested elsewhere in the literature\cite{lherbier2013electronic, Pedersen-SCdoping} for nitrogen, would enhance the features discussed in this work due to the scaling of the effective mass term with scatterer strength.

\begin{figure}[t]
	\centering
	\includegraphics[width =0.45\textwidth]{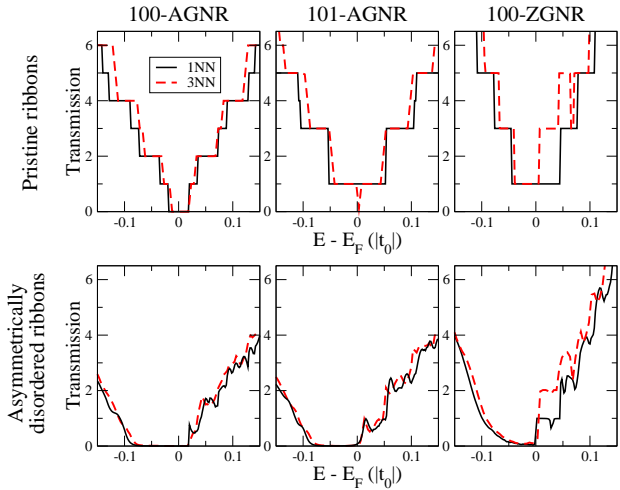}
	\caption{Transmissions for pristine (top) and asymmetrically disordered (bottom) ribbons using both 1NN (solid, black curves) and 3NN (red, dashed curves) models. The 1NN results for the 101-AGNR and 100-ZGNR are reproduced from the main text, whereas the 100-AGNR case represents an initially semiconducting ribbon within the 1NN model. }
	\label{fig_3NN}
\end{figure}

For AGNRs we note that the 1NN model captures all the main features, with the exception of the previously reported small band gap for pristine 101-AGNRs. 
We also note the band gap opening induced by asymmetric disorder occurs regardless of the metallic or semiconducting nature of the corresponding pristine ribbon.
The higher transmission values for pristine ZGNRs at low electron-side energies are due to the zero-energy peak no longer remaining disperionless within the 3NN model. This has been reported previously in the literature.\cite{Chang3NN_pores}
We note that the key result discussed in our paper, namely the band gap opening or transmission suppression at low hole-side energies in asymmetrically doped systems, are perfectly described by the 1NN model.


%

\end{document}